\documentstyle[12pt,aaspp4]{article}

\begin{document}

\title{Linear Polarization and Proper Motion in The Afterglow of Beamed GRBs}
\author{Re'em Sari}
\affil{Theoretical Astrophysics, 130-33, California Institute of Technology,
Pasadena, CA 91125 \newline
Institute for Theoretical Physics, University of California, Santa-Barbara,
CA 93106-4030}

\begin{abstract}
We investigate the polarization and proper motion expected in a beamed
Gamma-Ray Burst's ejecta. We find that even if the magnetic field has well
defined orientation relative to the direction of motion of the shock, the
polarization is not likely to exceed $20\%$. Taking into account the
dynamics of beamed ejecta we find that the polarization rises and decays
with peak around the jet break time (when the Lorentz factor of the flow is
comparable to the initial opening angle of the jet). Interestingly, we find
that when the offset of the observer from the center of the beam is large
enough, the polarization as function of time has three peaks, and the 
polarization direction of the middle peak is rotated by $90^\circ$ relative
to the two other peaks. 
We also show that some proper motion is expected, peaking around the
jet break time. Detection of both proper motion and the direction of
polarization can determine which component of the magnetic field is the
dominant.
\end{abstract}

\section{Introduction}

For the first time, polarization was measured from an optical afterglow, in
the case of GRB 990510 (Covino et. al. 1999, Wijers et. al. 1999). 
The polarization was
relatively small, 1.7\% and was measured about 0.77d after the burst. A Large
amount of polarization is usually considered as the smoking gun of
synchrotron emission. Awhile before this detection, some prediction
regarding polarization were put forward by Gruzinov and Waxman (1999) and
Medvedev and Loeb (1999). They assumed spherical emission, which by symmetry
give no net polarization except from random fluctuations. The 
polarization from those fluctuations depend on the size of the coherent
magnetic field regimes, the smaller they are the more chances of averaging
down to zero. The prediction were up to 10\% by Gruzinov and Waxman and of
order 1\% by Medvedev and Loeb.

However, the optical light curve of GRB 990510, for which polarization was
measured, showed a strong break into a steeper decline at about 1.5d (Stanek
et. al. 1999, Harrison et. al. 1999). Though spherical models predict
several breaks in the spectra and light curve of afterglow (see Sari, Piran
and Narayan 1998), these breaks are considerably smaller. The break seen in
GRB 990510 is therefore interpreted as a result of a beamed ejecta that
begins to spread sideways (Rhoads 1999, Panaitescu and M\'esz\'aros 1998, Sari,
Piran and Halpern 1999). Motivated by the measurement of the polarization,
and the interpretation of a beamed emission, Gruzinov (1999) recalculated
the polarization expected from a beamed ejecta and concluded that if the
magnetic field perpendicular to the shock front is significantly different
from the magnetic field parallel to the shock front then the amount of
polarization may be as large as the maximal synchrotron polarization, i.e.
about 60\%.

Here we reanalyze the expected polarization from a beamed ejecta. We take
an approach similar to that of Gruzinov (1999) where we assume that the
magnetic field is not completely isotropic behind the shock, with the
component parallel to the shock front significantly different from the other
components. The discussion here is different from that of Gruzinov in two
key points. First the polarization of the emission of powerlaw distribution
of electron, as observed in a given frequency is estimated (rather than the
frequency integrated polarization). Second, and more important, a more
realistic geometric setup for the afterglow emission is considered rather
than a point-like emitter. We take into account the evolution of the
relativistic jet and derive the time dependent polarization.

We denote by $\Pi _{0}$ the polarization from a small region where the 
magnetic field has a given orientation.  
In section \ref{secpoint}, the polarization $\Pi _{0}$ is
averaged over the possible orientation of the magnetic field. This is done
assuming that the magnetic field is entangled over a point-like region, 
in which the direction towards the observer is approximately constant. 
In section \ref{secbeam}, we integrate
over the entire emitting region in a beamed ejecta geometry, and obtain the
observed polarization. By producing these steps in that order one assumes
that the magnetic field is randomized on scales which are point-like, much
smaller than the overall size of the emitting region. As pointed out by
Gruzinov and Waxman (1999), this is true even if the magnetic field
coherent length grows in the speed of light in the fluid local frame. In
section \ref{secproper} we discuss the proper motion associated with a
beamed ejecta.

\section{\label{secpoint}Polarization from a point-like emitting region.}

\begin{figure}[tbp]
\begin{center}
\epsscale{.9} \plotone{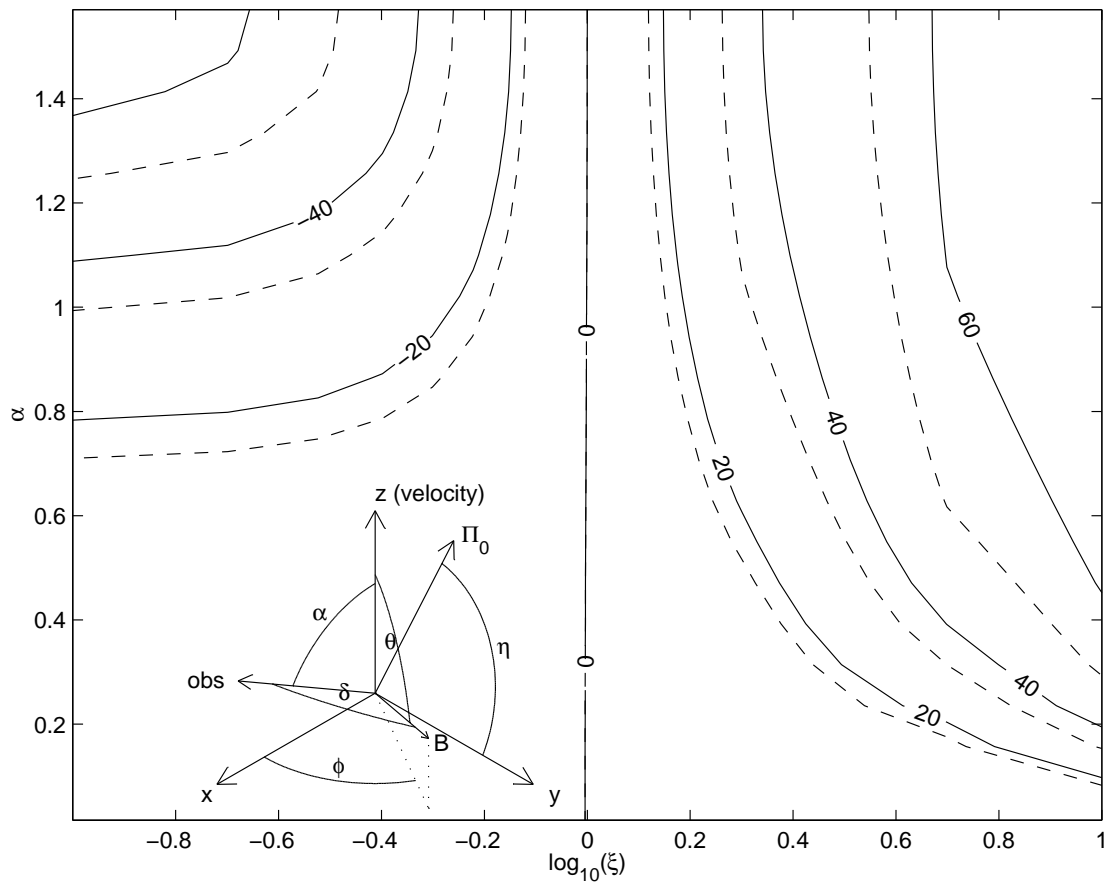}
\end{center}
\caption{ The percentage of polarization $\Pi_p$ from a point-like emitting 
region, 
after averaging over the possible orientation of the magnetic field, 
as function of the inclination of the observer relative to the 
preferred direction $\alpha$
and the ration between the two components of the magnetic field $\xi$. 
The insert shows the geometry of the calculation, which averages over
$\theta$ and $\phi$ for a given $\alpha$.
}
\label{point}
\end{figure}

At any point in the shock front there is a preferred direction, 
the radial direction, in which the
fluid moves. We call this the parallel direction and choose the $z$%
-direction of the fluid local frame coordinate system to be in that
direction. The two perpendicular directions $(x,y)$ are assumed equivalent,
i.e., the system is isotropic in the plane perpendicular to the direction of
motion. We chose the $x$-direction to be in the plane that contains the $z$%
-direction and the direction towards the observer $\widehat{n}$. Suppose now
that the magnetic field has spherical coordinates $(\theta ,\varphi )$ in
that frame (see insert in figure \ref{point}). A quite general description of the
distribution of the magnetic field in such anisotropic system would be to
allow different values of the magnetic field as function of the inclination
from the preferred direction $B=B(\theta )$ as well as a probability
function for the magnetic field to be in each given inclination $f(\theta )$.

The relevant component of the magnetic field is that perpendicular to the
observer i.e. $B\sin (\delta )$, where $\delta $ is the angle between the
direction of the magnetic field and the observer. This will produce
polarization $\Pi _{0}$ in the direction perpendicular both to the observer
and to the magnetic field i.e. in the direction $\hat{n}\times \hat{B}$.
However, this polarization should be averaged due to contributions from
magnetic fields oriented differently. By our assumption of isotropy in the $%
(x,y)$ direction, the polarization of radiation emitted from a point-like
region (after averaging on magnetic field orientation) must be in the
direction perpendicular to the $z$-axis and to the observer, i.e., in the $%
\hat{y}$ direction. The contribution $\Pi _{0}$ from a single orientation
magnetic field, must therefore be multiplied by $\cos 2\eta $ where $\eta $
is the angle between $\hat{y}$ and $\hat{n}\times \hat{B}$. By doing so,
positive total polarization would indicate polarization along the $\hat{y}$
direction while negative polarization would indicate polarization along the
direction perpendicular to $\hat{y}$ and to the observer. Assume now that
the emission is proportional to some power of the magnetic field $%
B^{\epsilon }.$ The total polarization from a point-like region is then
\[
\Pi _{p}=\Pi _{0}\frac{\int \cos (2\eta )[B(\theta )\sin \delta ]^{\epsilon
}f(\theta )\sin \theta d\varphi d\theta }{\int [B(\theta )\sin \delta
]^{\epsilon }f(\theta )\sin \theta d\varphi d\theta }. 
\]
For a powerlaw distribution of electrons we have $\Pi _{0}=(p+1)/(p+7/3)$
and $\epsilon=(p+1)/2$,
where $p$ is the electron powerlaw index, usually in the range of $p=2$ to $%
p=2.5$. Reasonable values are therefore $\Pi _{0}\sim 70\%$ 
and $1.5<\epsilon<1.75$. Cooling may increase the effective $p$ by $1/2$.
The angles $\delta$ and $\eta$ are given by 
\[
\cos \delta =\cos \alpha \cos \theta +\sin \alpha \sin \theta \cos \varphi , 
\]
\[
\cos \eta =(\sin \alpha \cos \theta -\cos \alpha \sin \theta \cos \phi
)/\sin \delta . 
\]

For frequency integrated polarization, the emission is proportional to the
square of the magnetic field, $\epsilon =2$, and the integration can be
easily done. We obtain 
\[
\Pi _{0}\sin ^{2}\alpha \frac{<B_{\parallel }^{2}>-<B_{\perp }^{2}>/2}{\sin
^{2}\alpha <B_{\parallel }^{2}>+(1+\cos ^{2}\alpha )<B_{\perp }^{2}>/2}. 
\]
This is identical to the expression of Gruzinov (1999). As we remarked
above, the relevant values of $\epsilon$ are probably below 2, and 
the integration is less simple. The results now
depends on higher moments of $B(\theta )$ and $f(\theta )$, rather than
simply through $<B_{\parallel }^{2}>$ and $<B_{\perp }^{2}>$. One
realization of anisotropic magnetic field can be obtained from an isotropic
magnetic field in which the component in the parallel direction was
multiplied by some factor $\xi $. In the notation above this translates to 
\[
B(\theta )\propto \left( \sin ^{2}\theta +\cos ^{2}\theta /\xi ^{2}\right)
^{-1/2}, \ \ \  
f(\theta )\propto \left( \sin ^{2}\theta +\cos ^{2}\theta /\xi ^{2}\right)
^{-3/2}. 
\]
Figure \ref{point} shows contours of the polarization obtained from a single
emission point after averaging over all possible direction of the magnetic
field relative to $\Pi_0$, 
as function of the inclination angle $\alpha $
and the anisotropy $\xi$, both in the
case of $\epsilon =1.5$ and the analytic case $\epsilon =2$. The $\epsilon
=2 $ cases gives higher polarization than lower values of $\epsilon$.
However, it is evident that the differences in polarization
for the two values of $\epsilon $ is not large and are mostly less than $%
10\% $. Given the much higher uncertainties, such as the anisotropy $\xi$
and the uncertainty in the geometry when averaging over the emitting regions
(see the next section), one can use the analytic result even though it uses $%
\epsilon =2>1.75$. The following properties seems to be general: if $\xi \gg
1$ ($B_{\parallel }\gg B_{\perp }$) the polarization is $\Pi _{p}\cong \Pi
_{0}$ quite independent on the exact value of $\alpha $ and $\epsilon $. If,
on the other hand, $\xi \ll 1$ ($B_{\parallel }\ll B_{\perp }$) as suggested
by Medvedev and Loeb (1999) then the polarization is small for small values
of $\sin ^{2}\alpha $. In the following section we shall assume the favorite
conditions in which the polarization from a point like emitter is $\Pi
_{p}\cong \Pi _{0}\cong 70\%$.

\section{\label{secbeam}Polarization from a beamed relativistic ejecta.}

The exact calculation of the expected polarization requires the knowledge of
the exact hydrodynamics of the evolution of a beamed ejecta. There is no
detailed description of that for now, however, several key features are
understood. At first, the ejecta behaves like a spherical one as it has no
time to spread laterally. This stage lasts as long as $\gamma \gg \theta
_{0}^{-1}$ where $\gamma $ is the bulk Lorentz factor of the jet and $\theta
_{0}$ is its opening angle. During this stage the emission also looks
spherical, since the observer is only able to see a small fraction of the
jet surface, of the order of $\gamma ^{-1}\ll \theta _{0}.$ The emission in
this stage is mostly from a ring at high frequencies (above the peak
synchrotron frequency) and from an almost uniform disk below the peak
frequency and especially uniform below the self absorption frequency (Waxman
1998, Panaitescu and M\'esz\'aros 1998, Sari 1999, Granot, Piran and Sari
1999a,b). Since the emitting region has spherical symmetry around the
observer, the net expected polarization is zero, except perhaps for
fluctuations in the manner discussed by Gruzinov and Waxman and Medvedev and
Loeb. We ignore this kind of fluctuations in the rest of this paper and
focus on the average net polarization.

It is likely that the observer is not directed exactly at the center of the
jet. In this case once the viewing angle becomes large enough the observer
``feels'' the asymmetry and most of the emission comes from the direction
towards the center of the jet. At this stage some net polarization is
expected. However, one should not expect the maximal linear polarization
allowed by synchrotron radiation as the emission has angular extent $%
1/\gamma $ and therefore a considerable averaging will take place. The time
when the edge effects become visible is comparable to the time when the jet
begins to spread. Later in time the angular extent of the jet increases,
while the offset of the observer from the center of the jet is, of course,
fixed in time. The observer therefore becomes more and more in the center of
the ejecta and the system, once again, approaches cylindrical symmetry. The
amount of polarization is expected to fade.

When the emission is from a ring centered around the observer, and assuming
that the dominant component of the magnetic field is perpendicular to the
shock then the north and south quarters of the ring produce polarization in
the north-south direction while the west and east quarters give rise to
west-east polarization (The opposite is true if the parallel magnetic field
is the dominant one). The total being a zero net polarization. If part of
the ring is missing (due to the finite extent of the jet) say a small part
on the east direction, then the net polarization would be in the south-north
direction. If it is a big fraction of the ring that is missing, say the east
part as well as the north and south parts then the polarization would be
east-west. As discussed above, the part of the ring that is missing is
initially growing, reaching a maximum around the time when the jet begins to
spread, and then decreases again. If, at the maximum, a large part of the
ring is missing (or radiates less efficiently) then the direction of
polarization is expected to change by $90{{}^{\circ }}$! This kind of
behavior is quite unique to the geometric setup of beamed GRBs. A detection
of such a feature is therefore a very strong support both to the synchrotron
radiation as well as the geometric structure of the jet and its evolution.
Some possible examples of this behavior is given in the toy model below.

We suggest a toy model in order to get a better filling of the possible
observable effects and a very rough estimate of the maximal polarization.
The toy model is built on the following assumptions: {\bf 1. }The line of
sight to the observer is always crossing the jet, i.e. the angular offset
between the observer and the center of the jet is smaller than the initial
angular extent of the jet $\theta _{0}$. {\bf 2. }The viewable region is a
thin ring of radius $\gamma ^{-1}$ centered around the line of sight to the
observer. The width of the ring is taken to be 30\% of it radius. {\bf 3. }%
The jet spans an angular size given by its initial size $\theta _{0}$ as
long as $\gamma \ge \theta _{0}^{-1}$ and after that expands with angular
size of $\gamma ^{-1}$, i.e., $\theta (t)=\max [\gamma ^{-1},\theta _{0}]$. 
{\bf 4. }The portions of the viewable region (defined in {\bf 2}) that
overlaps the jet radiates uniformly, while the portion of the viewable
region that is outside of the jet is not emitting. {\bf 5. }The Lorentz
factor of the fluid is related to the observed time, $T$, by $\gamma
\propto T^{-3/8}$ as long as $\gamma \ge \theta _{0}^{-1}$ and by $\gamma
\propto T^{-1/2}$ after that.

Under these assumptions, the evolution of the polarization as a function of
time depends only on the initial offset between the line of sight to the
observer and center of the jet measured in units of the jet's initial
angular size $q\equiv \theta _{offset}/\theta _{0}.$ Jets with $q=0$ are
centered exactly at the observer, while with $q=1$ the observer is located
exactly at the edge of the jet. Values of $q>1$ are excluded by our first
assumption. This reflects the fact that in such cases the GRB itself will be
hardly seen.

\begin{figure}[tbp]
\begin{center}
\epsscale{.9} \plotone{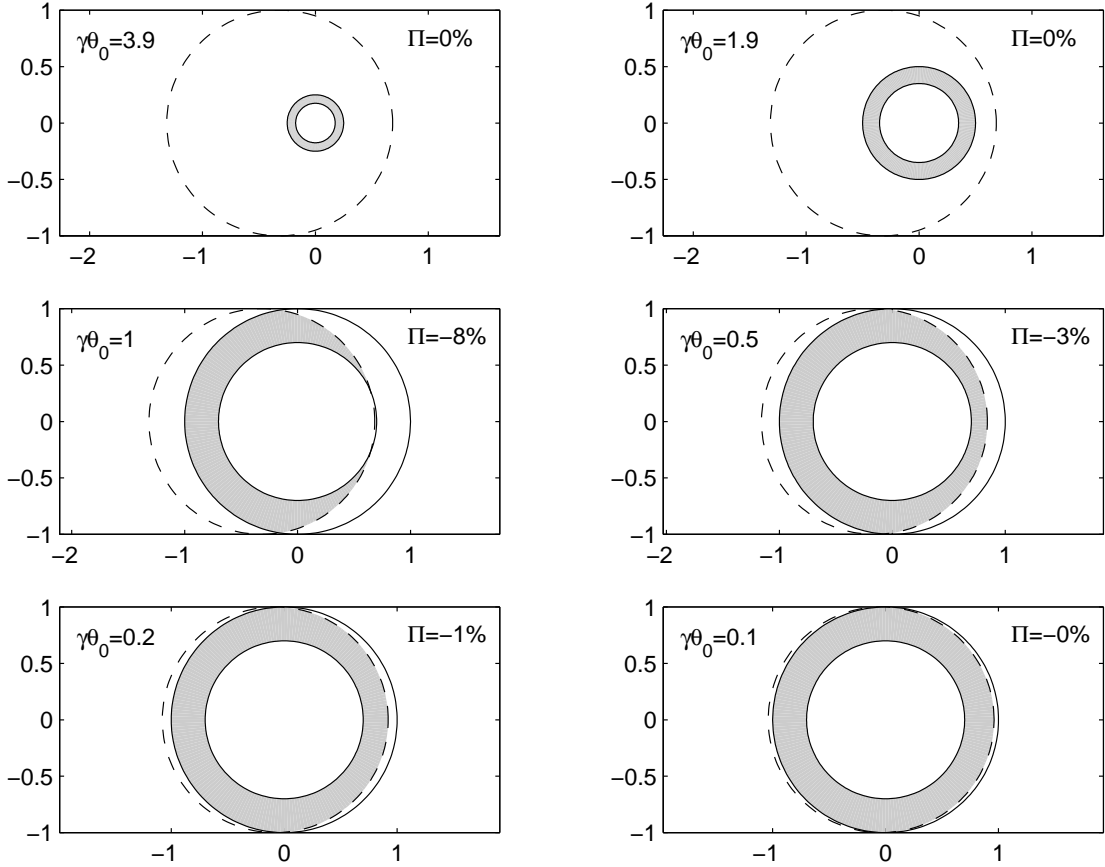}
\end{center}
\caption{ The emitting region at several times in the toy model, the case of
a relatively small offset of $q=0.32$. Dash line marks the physical extent 
of the jet while solid lines give the viewable region.
The gray shaded region is where the radiation is coming from.
On each frame, the percentage of linear 
polarization is given on the top right and the initial size of the jet 
relative to $1/\gamma$ is given on the left. 
The frames are scaled so that the size of the jet is unity.
}
\label{fig1a}
\end{figure}

\begin{figure}[tbp]
\begin{center}
\epsscale{.9} \plotone{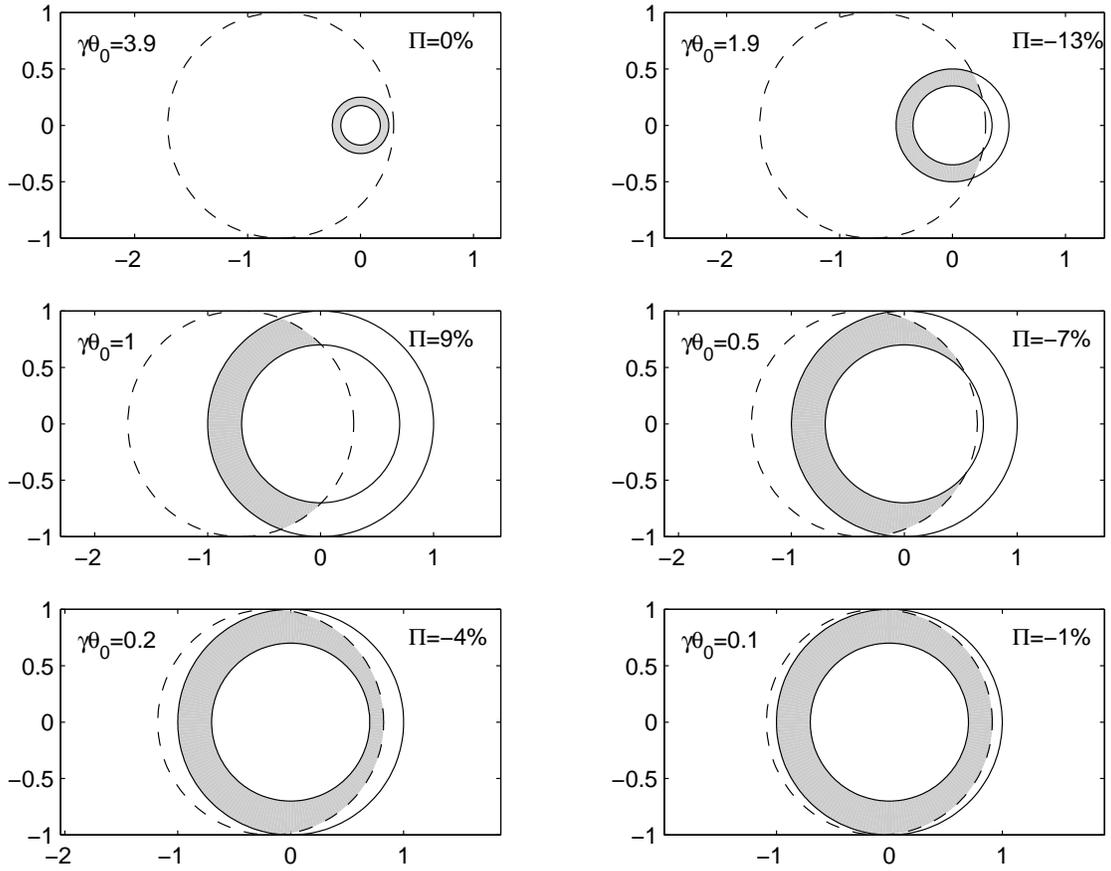}
\end{center}
\caption{ Same as figure \ref{fig1a} for a relatively large offset of $q=0.95$. }
\label{fig1b}
\end{figure}

\begin{figure}[tbp]
\begin{center}
\epsscale{.9} \plotone{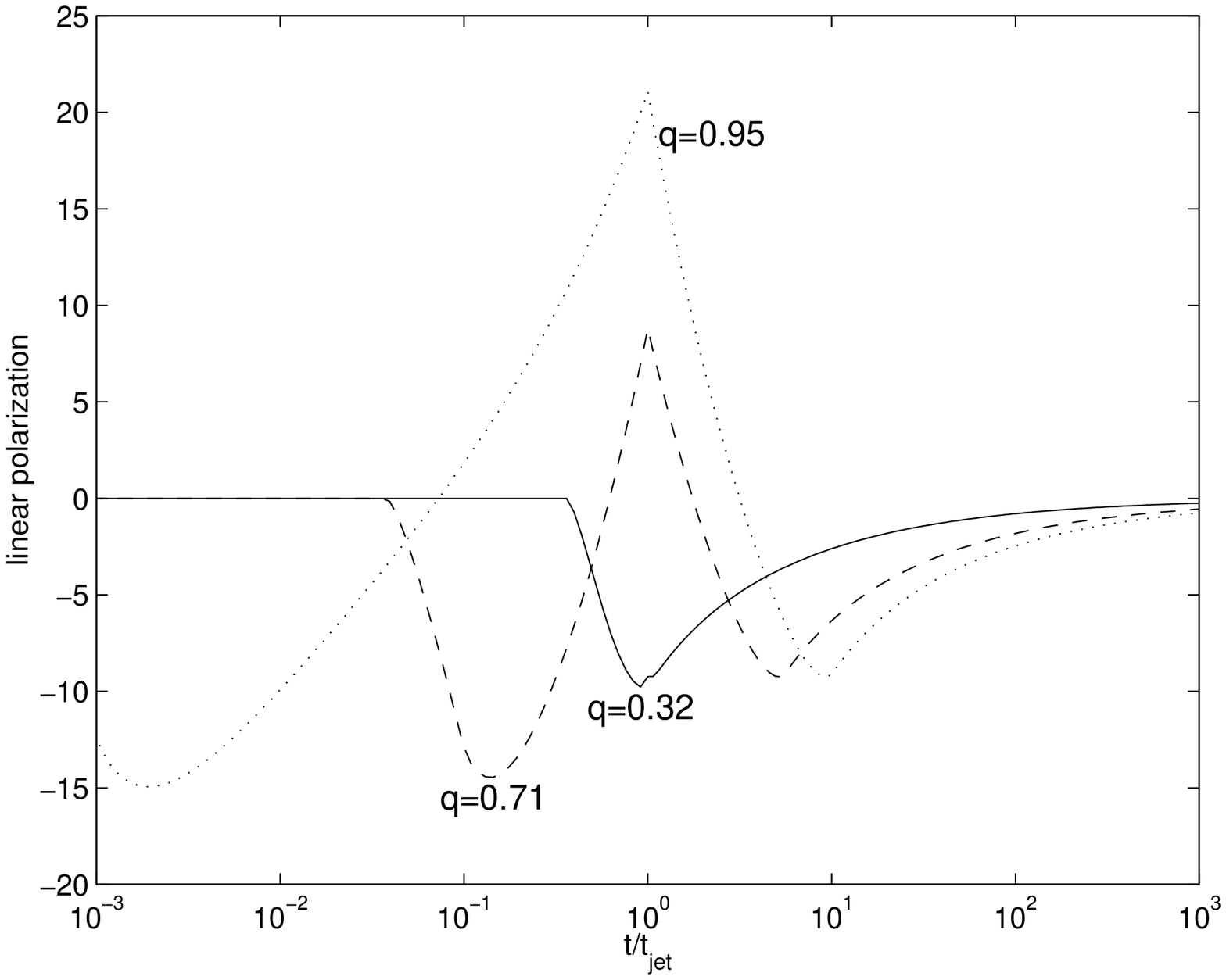}
\end{center}
\caption{ The polarization as function of time for three values of $%
q=0.32,0.71,0.95$. We assume here that the polarization from a single patch
is the maximal one, $\Pi_p=70\%$. }
\label{fig2}
\end{figure}

Figure \ref{fig1a} and figure \ref{fig1b} display the emission geometry for
the two radical values $q=\sqrt{0.1}\cong 0.32$ and $q=\sqrt{0.9}\cong 0.95$
where $80\%$ of the cases are. Figure \ref{fig2} summarizes the polarization
evolution for these two extreme values of $q$ as well as the median value $%
q=0.71$.

\section{\label{secproper}Proper Motion}

A straight forward consequence of the discussion and the toy model of the
previous section is proper motion of the source that peaks around the time
of the jet spreading. Since initially the source is centered around the
observer due to relativistic beaming, the centroid of the emitting ring is
fixed in time. Once the Lorentz factor becomes comparable to one over the
opening angle of the jet, the symmetry breaks and more emission comes from
the side directed towards the center of the jet. This results in proper
motion. 


The most extreme change in the angular position can be estimated by
$q\theta _{0}R/D$, where $R$ is the emission radius and $D$ is the (angular)
distance to the observer. 
Even with favorite parameters of $\theta _{0}=0.2$, $q=1$, $R=10^{18}$cm
and a distance of $D\sim 2\times 10^{27}$cm ($z\cong 0.2$) we get angular
displacement of order of $\sim 10^{-10}\cong 20\mu $arcsec, which is about
the current observational capabilities with VLBI. A more detailed time 
dependent calculation could be done by finding the centroid of the 
gray regions in figure \ref{fig1a} and figure \ref{fig1b}.

%

Since the proper motion is towards the
center of the jet, it must be either perpendicular or parallel
to that of the polarization. If the direction of the motion is parallel
(perpendicular) to the direction of the polarization (during the first peak
in case that the polarization changes its direction) then the dominant
component of the magnetic field is the parallel (perpendicular) component.
Combination of detection of polarization and proper motion enables as to
determine the orientation of the magnetic field behind the shock.

\section{Discussion}

We have estimated the polarization expected in the case where the magnetic
field is not completely randomized behind the relativistic shock. High
polarization is expected if the magnetic field is significantly different in
the parallel and perpendicular directions. We find that the polarization due
to powerlaw distribution of electrons is only slightly smaller than the
frequency integrated polarization. However, averaging over the whole
emission site has a dramatic effect on the total polarization. It completely
destroys the polarization at early and late times and polarization is 
expected only around the jet break
time. Even under the extreme conditions of the toy model, the polarization
is unlikely to get to its maximal value of $\sim 70\%$. The maximal value we
get is around $20\%$. 

A striking and quite unique outcome of our model is that the polarization
change direction by $90{{}^{\circ }}$ around the jet spreading time for
cases where the observer is not very close to the center of the jet. It
rises once in a given direction decays to zero and rises again in a
direction different by $90{{}^{\circ }}$ vanishes again and finally rises in
the original direction and slowly decays to zero. Within our toy model, most
beamed ejecta are expected to change the direction of their polarization in
the above manner.

Beamed GRBs are subject to proper motion of the centroid of the
emission region, mostly around the jet spreading time. On early time, the
emission is centered around the observer and there is therefore no motion.
Around the jet spreading time, the position angle should change by a small
amount, of order of a few $\mu$arcsec marginally detectable by current 
instruments. If proper motion
is detected, it would be towards the center of the physical jet. This is
either perpendicular or parallel to the polarization direction. It
will tell as which component of the magnetic field is larger.

Ghisellini and Lazzati (1999) have simultaneously and independently 
completed a similar work, and found two peaks for the polarization 
in the limit of $\xi=0$ and a non spreading jet.
The jet spreading effect, which is taken into account
in this letter, brings back the symmetry at late times and
results in a third polarization peak, in the same direction as the
first peak. The spreading also destroys the second peak if the offset is
small enough.

\acknowledgments
I thank Roger Blandford, Tsvi Piran, Eric Blackman, Sterl Phinney and Dail
Frale for useful discussions. The author is supported by the Sherman
Fairchild foundation. This research was supported in part by the National
Science Foundation under Grant No. PHY94-07194.


\begin{references}
\reference{covino} Covino et. al. 1999 A\&A in press, astro-ph/9906319. 
\reference{GPS98}Granot, J., Piran, T., \& Sari, R., 1999a, ApJ, 513, 679.
\reference{jgranot98b} Granot, J., Piran, T., \& Sari, R., 1999,
astro-ph/9808007.
\reference{gw99} Gruzinov, A. and Waxman, E. 1999 ApJ, 511, 852.
\reference{g99} Gruzinov, A. 1999 ApJL submitted, astro-ph/9905276.
\reference{harrison} Harrison F. A. et. a. 1999, astro-ph/995306.
\reference{harrison} Ghisellini, G. \& Lazzati D., 1999, astro-ph/9906471.
\reference{medvedev} Medvedev, M. V. and Loeb A. 1999, astro-ph/9904363.
\reference{Meszaros97}Panaitescu, A., \& M\'esz\'aros, P., 1998,  
             astro-ph/980616.
\reference{Rhodas99}Rhoads, J.E., 1999, Astro-ph/9903399.
\reference{Sari97b} Sari, R., 1998, ApJL, 494, L49.
\reference{SNP} Sari, R., Piran, T. \& Narayan, R., 1998, ApJ, 497, L17.
\reference{SPH} Sari, R., Piran, T. \& Halpern, J., 1999, ApJ, 519, L17.
\reference{SNP} Stanek et. al, 1999, astro-ph/995304.
\reference{Waxman97c}Waxman, E., 1997, ApJL, 491, L19.
\reference{Wijers} Wijers, R.A.M.J., et. al. 1999, astro-ph/9906346.

\end{references}
\end{document}